# Optical Magnetism: from Red to Blue

Wenshan Cai, Uday K. Chettiar, Hsiao-Kuan Yuan, Vashista de Silva, Alexander V. Kildishev, Vladimir P. Drachev, and Vladimir M. Shalaev

*Birck Nanotechnology Center, Purdue University, West Lafayette, Indiana 47907, USA*

**Abstract:** A family of coupled nano-strips with varying dimensions is fabricated to obtain optical magnetic responses across the whole visible spectrum, from red to blue. The proposed approach provides one with a universal building block and a general recipe for producing controllable optical magnetism for various practical implementations.

Light, as an electromagnetic wave with both electric and magnetic components, interacts with natural materials in a "one-handed" manner. Almost all conventional optical phenomena result from effects associated with the electric component of light, while the magnetic component barely plays a role. A partial explanation may be given within the regime of classical atomic electrodynamics, which shows that the effect of light on the magnetic polarizability is roughly $\alpha^2$ ($< 10^{-4}$) weaker than its electric counterpart, with $\alpha$ being the fine structure constant [1]. There are two additional facts that further prohibit any natural optical magnetism. First, the selection rule for non-relativistic magnetic dipole transitions determines that such transitions are allowed only between states with the same spatial parts of wavefunctions [2], and the energy difference between such two states is much smaller than the usual photon energy, which is on the order of 1 eV. Second, there are no existing magnetic monopoles, therefore one cannot produce a magnetic plasma as can be accomplished with electrons in metals at optical frequencies. Consequently, as instructed by many textbook writers, including Lev Landau, "there is certainly no meaning in using the magnetic susceptibility from optical frequencies onwards, and in discussion of such phenomena we must put $\mu = 1$ [3]."

This situation has started to change in recent years since Pendry suggested using artificially structured "atoms" of a metamaterial to mimic magnetism at higher frequencies [4]. Magnetic responses based on split-ring resonators (SRRs) and their analogues have been reported from C-band microwave frequencies up to the optical wavelength of 800 nm [5-9][*]. At visible wavelengths, however, other structures like coupled nano-rods [10] or nano-strips [11-13] are preferred because there are intrinsic limits to scaling SRR sizes in order to exhibit a magnetic response in the optical range [14,15][**]. We note that any controllable optical magnetic responses, whether they have a positive or negative permeability, are important for various implementations such as negative refraction [18-20], subwavelength waveguides and antennas [21,22], spectral selective filters [23], total external reflection [24], and electromagnetic

---

[*] Here we use frequency-dispersive permittivity $\varepsilon(\omega)$ and permeability $\mu(\omega)$ to characterize the electromagnetic response of a metamaterial. There is also an alternative description based on the generalized, spatially dispersive permittivity tensor $\tilde{\varepsilon}(\omega,\vec{k})$ to describe both electric and magnetic responses without using permeability $\mu$.

[**] Herein we do not consider the reported magnetism at the green light by coupled nano-pillars [16] because subsequent studies by two independent groups show that the claim is unjustified [17].

cloaking devices [25,26].

In this work we study the general resonant properties of magnetic metamaterials consisting of arrays of paired thin silver strips. The magnetism in such a structure has been discussed theoretically [11,12] and was recently demonstrated experimentally at the very red end of the visible range [27]. Here we extend the studies to obtain magnetic responses across the whole visible spectrum. We create a family of paired-strip samples with varying geometries. The dependence of the magnetic resonance wavelength on the geometric parameters is studied both experimentally and theoretically.

Fig. 1a shows a cross-sectional schematic of the coupled nano-strip samples. A pair of thin silver strips with thickness $t$ is separated by an alumina spacer with thickness $d$ and refractive index $n_d \approx 1.62$. The whole sandwich stack is trapezoidally shaped with an average width of $w$ and a bottom width of $w_b$ to reflect the reality of fabrication. The thickness of each silver layer and the alumina spacer are $t = 35$ nm and $d = 40$ nm, respectively. These parameters are optimized values based on spatial harmonic analysis. For different samples, we vary the width $w$ of the strips to obtain optical magnetic resonances at a set of wavelengths. The periodicity $p$ changes accordingly, such that the overall coverage ratio (defined by the ratio of bottom width $w_b$ to the periodicity $p$) of each sample is roughly 50%. This ensures that the strengths of the magnetic resonances in different samples are comparable.

The samples were fabricated by electron beam lithography (EBL) techniques. Each sample is 160 μm × 160 μm in size. During electron beam deposition two thin alumina layers of 10 nm were added to the top and bottom of the Ag-$Al_2O_3$-Ag sandwich stacks for fabrication stability. The geometric parameters of the six magnetic nano-strip samples are given in Table 1. All six samples are on the same substrate and were fabricated simultaneously to ensure a fair comparison. Fig. 1b and 1c show the field-emission scanning electron microscope (FE-SEM) and atomic force microscope (AFM) images of a typical paired nano-strip structure.

It has been demonstrated that nano-strip structures exhibit both magnetic and electric resonances under TM illumination with the magnetic field polarized along the strips (see Fig. 1a for polarization definition). However, for TE polarization with the electric field aligned with the strips, the structure has no resonant effects. To qualitatively illustrate the resonance properties of the magnetic samples with different strip widths, we took microscopic images of the samples for two orthogonal polarizations using an advanced optical microscope (Nikon Eclipse E1000) with polarization control. The microscopic images of all the samples within the same microscope view are shown in Fig. 2. For the resonant TM polarization case (Fig. 2a & 2c), we observe distinct colors in different samples in both transmission mode and reflection mode, which indicate the different resonant frequencies in different samples. For the non-resonant TE polarization, however, the colors are the same for all samples. In this case the samples act as diluted metals with a behavior similar to perfect metals: more reflection and less transmission at longer wavelengths. This is why the non-resonant images appear blue in transmission mode (Fig. 2b) and red in reflection mode (Fig. 2d).

In order to test samples with distinct resonance properties, we measured the broadband transmission and reflection spectra of the nano-strip samples using a custom apparatus with polarization control. The transmission and reflection spectra were normalized to a bare substrate and a calibrated silver mirror, respectively. The collected spectra are shown in Fig. 3. As expected, we observe strong resonances from both transmission and reflection spectra for the TM polarization (Fig. 3a and 3c). For TE polarization, however, the spectra display a non-resonant wavelength dependence over a broad wavelength range (Fig. 3b and 3d). The slopes of the spectra in TE mode confirm our claim regarding diluted metal as discussed before and explain again why samples look blue in transmission mode (Fig. 2b) and red in reflection mode (Fig. 2d) for TE polarization. The six samples were fabricated with a range of strip widths from 50 nm (Sample A) to 127 nm (Sample F), and we obtained magnetic resonances occurring from 491 nm to 754 nm, covering the majority of the visible spectrum. The positions of the resonant wavelengths in TM mode move towards the blue when decreasing the width of the strips from Sample F to Sample A. This verifies the scaling property of the magnetic structure of coupled nano-strips.

The spectra for TM polarization exhibit multiple resonances and several important characteristic wavelengths. In Fig. 4 we plot the transmission, reflection and absorption (including diffractive scattering) spectra of a typical paired-strip sample (Sample E) under TM polarization with three characteristic wavelengths marked on the curves. The magnetic resonance around $\lambda_{res,M}$ results from an anti-symmetric current flow in the upper and lower strips. Therefore, a circular current is excited in the cross-section plane of the structure by the incident magnetic filed and gives rise to a magnetic dipole response. This magnetic resonance is the major feature that we are pursuing in such a coupled nano-strips structure. In addition to the anti-symmetric current mode, the strip pair also supports a symmetric current mode at around $\lambda_{res,E}$, which results in an electric resonance. These two resonances induce the two local minima in the transmission spectra and local maxima in reflection, as illustrated in Fig. 4. The absorption spectrum in Fig. 4 shows that extra absorption occurs near the two resonance wavelengths $\lambda_{res,M}$ and $\lambda_{res,E}$, which is natural for plasmonic resonances in metal-dielectric structures. The transmission spectrum also displays a sharp turn-back at a relatively shorter wavelength $\lambda_{diff}$, which indicates the diffraction threshold of the grating-like structure. For a one dimensional grating with a periodicity $p$, a diffraction channel is created whenever the wavelength $\lambda$ reaches below a diffraction threshold given by $\lambda_{diff,j} = n_s p / j$, where $j$ is an integer and $n_s$ is the refractive index of the grating substrate (in our case, $n_s = 1.52$ for the glass substrate) [29]. When the wavelength $\lambda$ falls below $\lambda_{diff,j}$, strong distortion in the transmission is present, which is usually attributed to Wood's anomaly [30]. In our experiments this threshold is observed for Samples D, E and F, whose first-order threshold $\lambda_{diff,1}$ is within the detection range of $\lambda > 400$ nm. The positions of $\lambda_{diff,1}$ obtained from Fig. 3a for the three samples agree extremely well with calculated values, exhibiting deviations of less than 1%. When a diffraction channel is opened at $\lambda < \lambda_{diff}$, substantial optical power transfers to diffractive scattering. This is also observed from the spectra in Fig. 4.

In addition to experimental characterization, the properties of the coupled-strip samples were investigated by numerical simulations with a commercial finite element package. The material properties of silver are taken from well-accepted experimental data [31], with the imaginary part of the permittivity serving as an adjustable parameter to reflect the actual imperfection of the metal quality in EBL fabrication. The detailed techniques in simulating such structures were published elsewhere [27]. The transmission, reflection and absorption spectra for a representative sample (Sample E) are plotted in Fig. 4 along with the experimental data. All of the features observed in the experimental spectra are reproduced remarkably well in our numerical simulations.

For practical designs and applications, it is desirable to have an analytical expression for the relation between the magnetic resonance wavelength $\lambda_{res}$ (identical to $\lambda_{res,M}$ previously) and the geometric parameters ($w$, $d$, $t$) of the paired-strip structures. Following the cavity model approach discussed in [32], we find that $\lambda_{res}$ satisfies the following equation:

$$\varepsilon'_m(\lambda_{res}) = 1 - \frac{n_d^2/t}{\sqrt{(\frac{\pi}{w})^2 - (\frac{2\pi n_d}{\lambda_{res}})^2}} \left( 1 + \tanh \frac{2/d}{\sqrt{(\frac{\pi}{w})^2 - (\frac{2\pi n_d}{\lambda_{res}})^2}} \right), \qquad (1)$$

where $\varepsilon'_m$ is the real part of the metal permittivity. Due to the trapezoidal shape of the actual paired strips, the actual resonant wavelength is longer than that predicted by Eq. (1) by an adjustment constant $\eta$. That is, all $\lambda_{res}$ in the above equation should be replaced by $\lambda_{res}/\eta$.

Eq. (1) has no analytical solution and can only be solved numerically. To obtain an explicit solution, we use the first-order approximations for the square root and the hyperbolic tangent functions in Eq. (1). The permittivity $\varepsilon'_m$ of silver is given by the Drude model with $\varepsilon'_m(\lambda) = 5 - \lambda^2/\lambda_p^2$, where $\lambda_p$ = 134.6 nm (corresponding to the plasma frequency of $\omega_p$ = 1.4×10$^{16}$ Hz or 9.2 eV [31]) is the plasma wavelength of silver. After some algebra we obtain the following approximate solution to Eq. (1):

$$\lambda_{res} = \sqrt{4 + \frac{n_d^2 w}{\pi t} + \frac{2n_d^2 w^2}{\pi^2 td}} \eta \lambda_p \qquad (2)$$

This solution verifies the intuitive conclusion that scaling down the width $w$ of the strips results in a shorter resonant wavelength $\lambda_{res}$. Moreover, there is a less intuitive conclusion that reducing the thickness $t$ of the metal strips tends to give $\lambda_{res}$ a red-shift, which has been observed in our recent experiments [27]. Another interesting observation from Eq. (1) is that, if the coupled strips are made of perfect metal with $\varepsilon'_m \to -\infty$, Eq. (1) would require that the square root in the denominator equal zero. Therefore, in this case the resonant wavelength $\lambda_{res}$ would only depend on one geometric parameter $w$ and would be independent of the thickness $t$ or separation $d$. This claim is consistent with the work on microwave magnetic media with paired metal wires [33], where the resonant wavelength solely depends on the length of the wires. Not surprisingly, for a perfect metal Eq. (1) gives $\lambda_{res}/n_d = 2w$, which is a natural solution for the basic mode of an electromagnetic cavity with a characteristic size of $w$.

In Fig. 5 we plot the dependence of the magnetic resonance wavelength $\lambda_{res}$ on the average width $w$ of the trapezoidal-shape paired strip samples both from experiments and analytical approaches. The experimental data for the relationship between $\lambda_{res}$ and $w$ is taken from Fig. 3 (a,c). In the analytical approach of Eq. (1) and the approximate solution given by Eq. (2), the adjusting constant $\eta$ is set to be 1.48. From Fig. 5 we can see that the results obtained from the Eq. (1) analytical method match the experimental data perfectly, and the approximate solution of Eq. (2) fits the experimental $\lambda_{res}(w)$ relation remarkably well. Therefore, equations (1) and (2) can be used as a general recipe for producing paired-strip magnetic metamaterials at any desired optical wavelengths. Fig. 5 also exhibits negligible saturation due to size-scaling, which indicates that such a structure is capable of producing optical magnetism at even shorter wavelengths.

As for the strength of the magnetic responses in those samples, we retrieved the effective permeability $\mu'$ of each sample around the magnetic resonance wavelength $\lambda_{res}$ using numerical simulations with the homogenization technique [11,34]. For each sample, the material properties and geometrical parameters used in the retrieval procedure guarantee good agreement between the simulated and experimental broadband spectra. The minimum values of permeability for the six coupled-strips samples are shown in Fig. 5. The permeability obtained in each sample is distinct from unity, as it would be in conventional optical materials, and is found to be -1.6 in Sample F for dark-red light of 750 nm and 0.5 in Sample A at the blue wavelength of less than 500nm. We note that for all the samples we study the magnetic resonance wavelength $\lambda_{res}$ is at least 5 times larger than the strip width $w$, and therefore the coupled-strips samples can indeed be regarded as two-dimensional metamaterials at the wavelengths of interest.

In summary, we have demonstrated a universal structure based on coupled nano-strips to create optical magnetic responses across the whole visible spectrum. The resonant properties of a family of such structures with varying dimensions were studied both experimentally and numerically. We obtained the dependence of the magnetic resonance wavelength on the geometric parameters from both experimental observations and an analytical model, which provides us with a general recipe for designing such magnetic metamaterials at any desired optical frequency. Additionally, it is possible to tune the magnitude of the effective permeability $\mu'$ by changing the coverage percentage of the stripes. Therefore, to a large extent, the coupled nano-strips structure can serve as a general building block for producing controllable optical magnetism for various practical implementations.

The authors would like to thank Jennie Sturgis of the Bindley Bioscience Center at Purdue University for her gracious help with optical microscopy images.

**Table 1**

Table 1. Geometric parameters of the magnetic nano-strip samples

| Sample # | Bottom Width $w_b$ | Average Width $w$ | Periodicity $p$ | Coverage % * |
|---|---|---|---|---|
| A | 95 | 50 | 191 | 0.50 |
| B | 118 | 69 | 218 | 0.54 |
| C | 127 | 83 | 245 | 0.52 |
| D | 143 | 98 | 273 | 0.52 |
| E | 164 | 118 | 300 | 0.55 |
| F | 173 | 127 | 300 | 0.58 |

* Cover ratio is calculated by the ratio of bottom width to the periodicity.

Figure 1

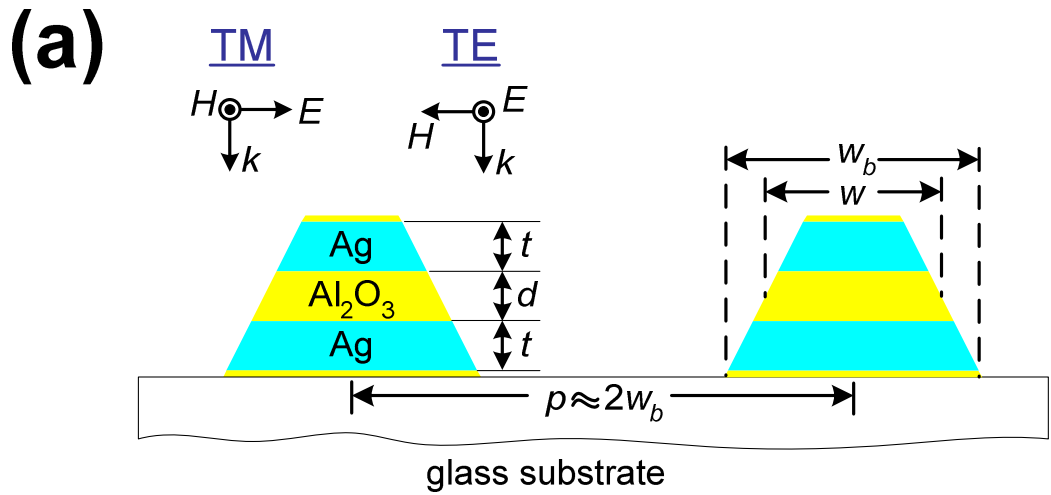

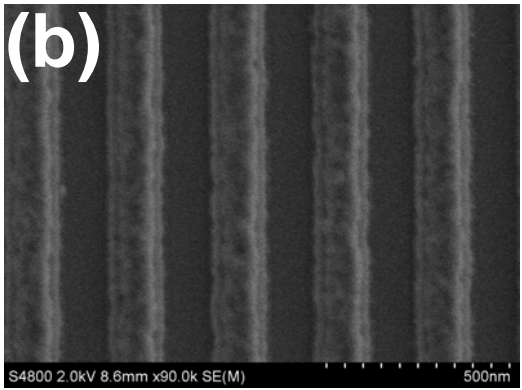
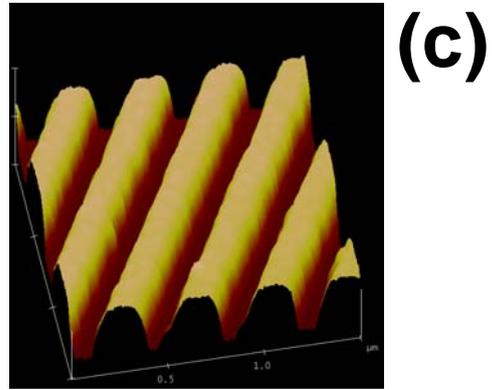

Figure 1: (a) The cross-sectional schematic of arrays of coupled nano-strips; (b) The FE-SEM image of a typical sample; (c) The AFM image of a typical sample. Pictures in panel (b) and (c) correspond to sample E in Table 1.

**Figure 2**

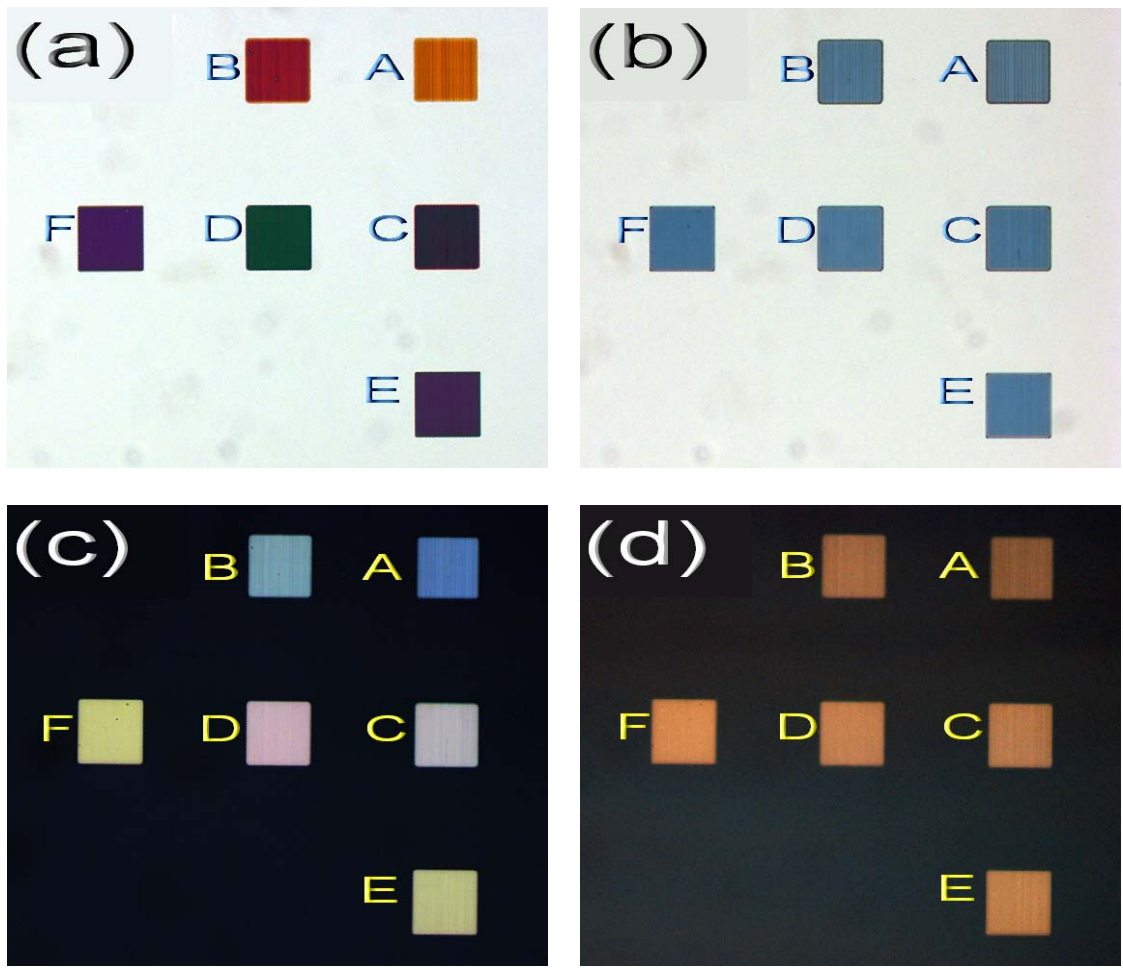

Figure 2: Optical microscopy images of the magnetic samples for two orthogonal polarizations. (a) Transmission mode with TM polarization; (b) Transmission mode with TE polarization; (c) Reflection mode with TM polarization; (d) Reflection mode with TE polarization. Letters A-F correspond to the sample naming in Table 1.

**Figure 3**

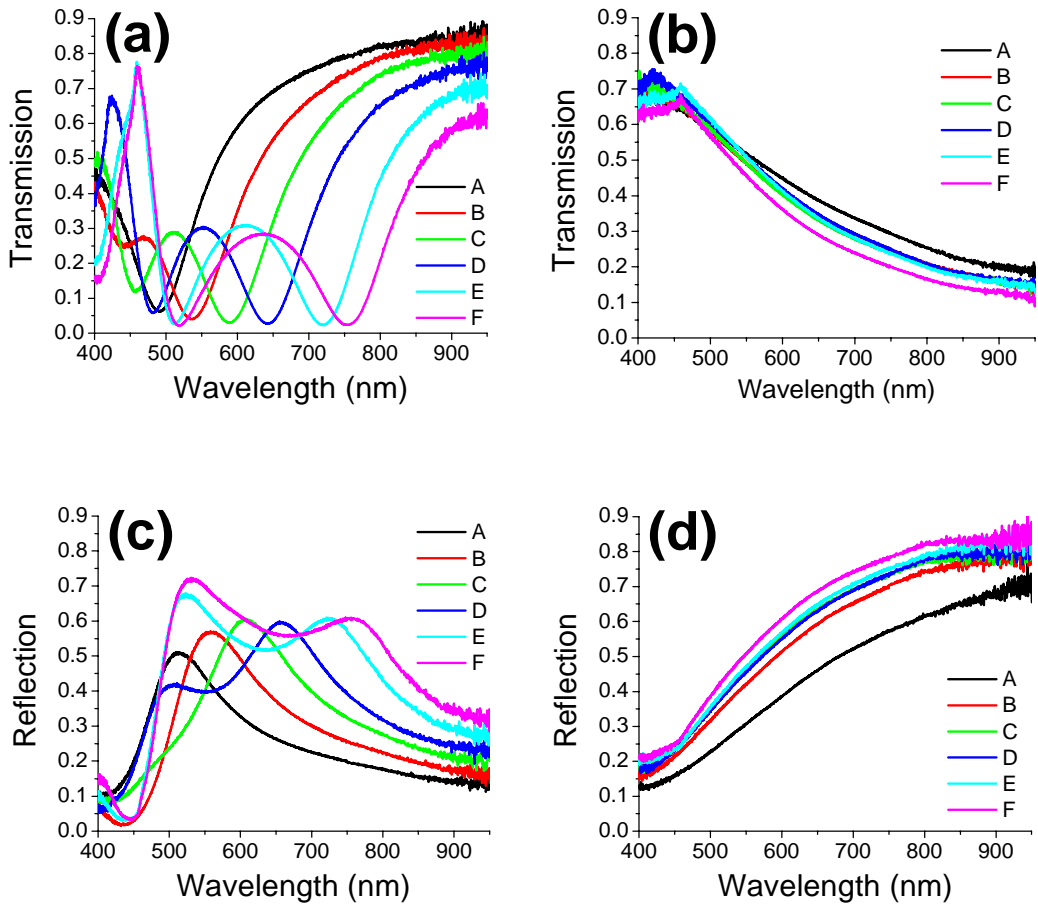

Figure 3: Transmission (T) and Reflection (R) spectra of the magnetic samples for two orthogonal polarizations. (a) T with TM polarization; (b) T with TE polarization; (c) R with TM polarization; (d) R with TE polarization. Letters A-F correspond to the sample naming in Table 1.

**Figure 4**

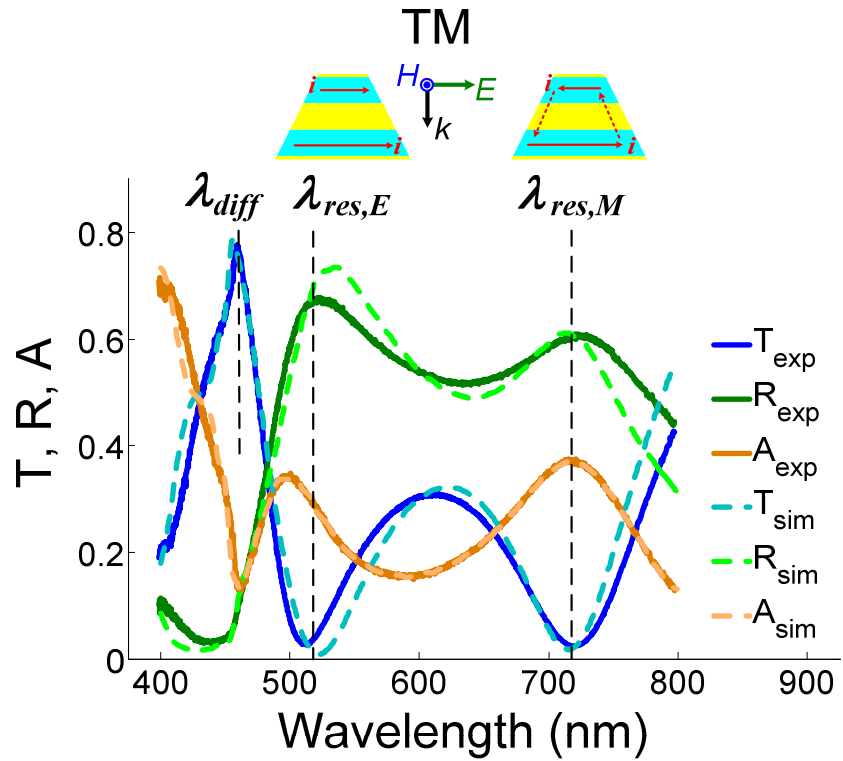

Figure. 4: Transmission (T), reflection (R) and absorption (A) (including diffractive scattering) spectra under TM polarization for a typical coupled nano-strip sample (Sample E) with three characteristic wavelengths marked. Solid lines show the experimental data, and dashed lines represent simulated results. The two cross-sectional schematics of the strip-pair illustrate the current modes at electric and magnetic resonances, respectively.

**Figure 5**

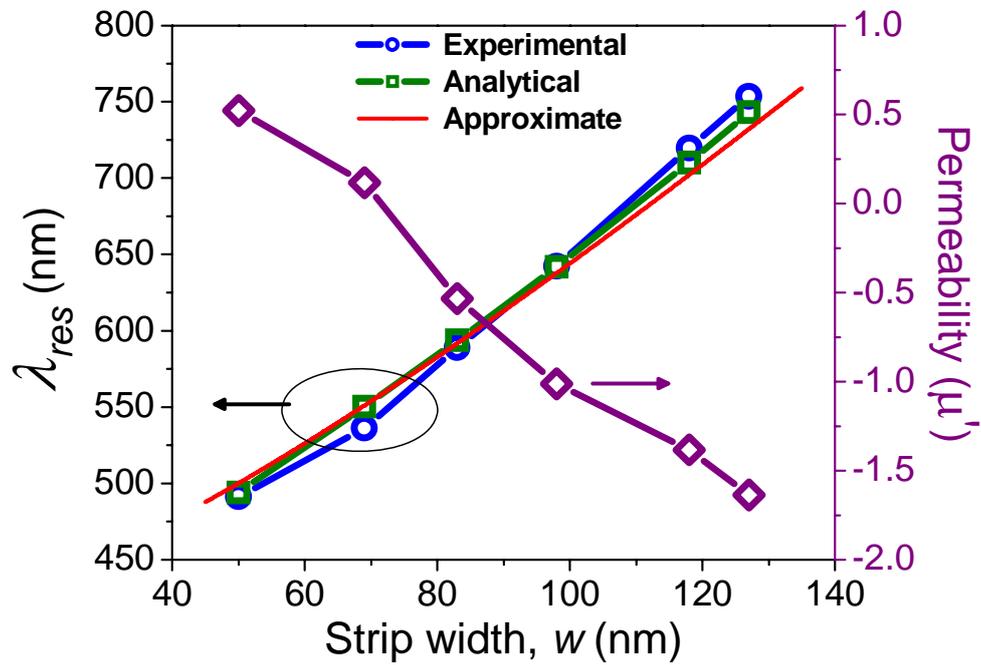

Figure 5: The dependence of the magnetic resonance wavelength $\lambda_{res}$ on the average width $w$ of the trapezoidal-shape paired strip samples, and the minimum values of permeability $\mu'$ for the six samples around $\lambda_{res}$. Circle: experimental data for $\lambda_{res}$ as a function of $w$ from Fig. 3 (a,c); Square: analytical $\lambda_{res}(w)$ relationship determined by Eq. (1); No Mark: approximate $\lambda_{res}(w)$ relationship given by Eq. (2); Diamond: retrieved minimum effective permeability for each sample.